\documentclass[twocolumn,preprintnumbers,amsmath,amssymb]{revtex4}
\usepackage{graphicx}% Include figure files
\usepackage{dcolumn}% Align table columns on decimal point
\usepackage{bm}% bold math
\newcommand{\etal}{\emph{et al.}~}
\newcommand{\be}{\begin{equation}}
\newcommand{\ee}{\end{equation}}
\newcommand{\bfig}{\begin{figure}}
\newcommand{\efig}{\end{figure}}
\newcommand{\incl}{\includegraphics}
\begin{document}      
\title{Theory of Asymmetric Tunneling in the cuprate superconductors} 
\author{P. W. Anderson and N. P. Ong}      
\affiliation{
Department of Physics, Princeton University, Princeton, N.J. 08544, U. S. A.
}

\date{\today}      
\begin{abstract}
We explain quantitatively, within the Gutzwiller-Resonating
Valence Bond theory, the puzzling observation of tunneling
conductivity between a metallic point and a cuprate high-$T_c$
superconductor which is markedly asymmetric between positive and
negative voltage biases. The asymmetric
part does not have a ``coherence peak" but does show structure due
to the gap. The fit to data is satisfactory within the
over-simplifications of the theory; in particular, it explains
the marked ``peak-dip-hump"  structure  observed on the hole side and
a number of other qualitative observations.  This asymmetry is
strong evidence for the projective nature of the ground state and hence
for ``$t$-$J$" physics.
\end{abstract}

\date{April 30, 2004}
%\begin{abstract}
%\end{abstract}


\maketitle
%\section{introduction}

In the conventional, BCS superconductors, the most complete and
convincing evidence for the phonon mechanism came from the
tunneling
spectrum--the tunneling conductivity as a function of junction
voltage.  The features in this spectrum were shown to be uniquely
caused by the anomalous self-energy (the ``gap") and gave
unequivocal evidence for its origin in the exchange of phonons.  It has
been
a disappointment that tunneling spectroscopy has so far given us no
such evidence for the cuprate superconductors.

One of the puzzling experimental features about the tunneling
in cuprates
is the fact that the tunneling conductivity is
markedly asymmetric as a function of voltage.  This was observed
even
in early crude attempts but became serious when vacuum tunneling
from
STM points
achieved clean results in good detailed agreement (except for this
fact) with the expected $d$-wave density of states~\cite{Pan,Pan2}.
It is particularly interesting when one realizes that asymmetries are
rare to non-existent in most metal-to-metal contacts, and are
predicted not to exist (except for slow, continuous variations of
tunneling
probabilities) within Fermi liquid theory.  The Gutzwiller-projected
mean-field-theory~\cite{Zhang}
     \cite{Anderson}
is not a Fermi liquid theory and can -- and does -- exhibit asymmetry.

The rarity of other examples of structure in tunneling-other
than the well-known Giaevar effect of the appearance of the
superconducting gap -- is a consequence of two remarkable theorems.
The first was proved by Harrison~\cite{Harrison}
essentially in order to explain why Giaevar
did not see band structure effects in normal metals.  Harrison
showed that the tunneling probability between two states, $\bf k$ in
metal A
and $\bf k'$ in metal B, evaluated in WKB approximation, contains a
factor
$v({\bf k})v({\bf k'})$ from the ``attempt frequency" -- where $v({\bf
k})$ is the velocity -- in addition to the WKB
integral.  (The theorem is actually more general than WKB, but
this will do).  This factor cancels against the density of states,
which is proportional to $1/v$. Although there may be prominent
density-of-states
anomalies in the band structure near to the Fermi level caused
by a Van Hove singularity, they will not show up strongly in
tunneling.

The second theorem is due to Schrieffer~\cite{Schrieffer}.
This is particularly useful in the BCS case,  but has some bearing in
the present one.
Schrieffer pointed out that in most situations the tunneling
probability is spread over a wide range of $\bf k$-values, so that one
must integrate the single-particle Green's function that appears in
the tunneling conductivity over the variable $\bf k$. This may be
converted into a
contour integral around the pole in the Green's function at the
quasiparticle energy, and simplifies the tunneling density-of-states
in the BCS case to

\be
N(E)=N(0){\rm Re}\left\lbrace{
{E}\over{\sqrt{E^2-\Delta^2(E)}}}\right\rbrace.
\label{eq:dos}
\ee

Here $\Delta(E)$ is the gap function evaluated at the energy of
the quasiparticle pole, where
$E^2 = [\epsilon_{\bf k}+\Sigma({\bf k},E)]^2 + \Delta({\bf k},E)^2$.

The result is not quite so clean in our case, where self-energies
can be assumed
to be $\bf k$- as well as $E$-dependent, but this modification seems
only to require a factor of $[1 + \partial\Sigma/\partial\epsilon_{\bf
k}]^{-1}$, which is
likely to vary rather smoothly, at least near the coherence peaks.

In the Gutzwiller mean-field-theory~\cite{Anderson}, we
start from the approximation that the ground state of the $t$-$J$  model
Hamiltonian is a
projected BCS product function, chosen by minimising the energy over all
such functions. The $t$-$J$  Hamiltonian is arrived at by a canonical
transformation~\cite{Gros} of the true Hamiltonian,
which presumably is essentially a Hubbard Hamiltonian:

\be
H_{tJ} = \mathrm{e}^{iS}H\mathrm{e}^{-iS} = \hat{P}T\hat{P} + J\sum_{<i,j>}{\bf S}_i \cdot
{\bf S}_j \
,
\label{eq:tj}
\ee
where $T$ is the kinetic energy and $\hat{P}$ the Gutzwiller projection
operator

\be
\hat{P} = \prod_i (1-n_{i\uparrow}n_{i\downarrow})\  .
\label{eq:proj}
\ee

The exchange term is not projected because it automatically
remains within the subspace defined by $\hat{P}$, and hence commutes with it.
The eigenstates of the $t$-$J$  Hamiltonian are necessarily
projected one-electron functions, so it is natural to approximate them
with
product functions.  Thus the ground state variational function is
\be
                  |f\rangle= \mathrm{e}^{iS}\hat{P} |\Phi\rangle
\label{eq:phi}
\ee
where $\Phi(\Delta,\mu)$ is the BCS product function
\be
|\Phi\rangle = \prod_{\bf k}
(u_{\bf k}+v_{\bf k}c^\dagger_{\bf k\uparrow}c^\dagger_{-k\downarrow}
|0\rangle,
\label{bcs}
\ee
and $u_{\bf k}$ and $v_{\bf k}$ are the variational parameters, determined by an
effective BCS Hamiltonian that gives us a gap equation as
discussed in Ref.~\cite{Zhang}.
This gap equation is actually the equation for the excitation
energies of Gutzwiller-projected excited-state wave functions
\be
|\Phi_{\bf k\sigma}\rangle=\mathrm{e}^{iS}\hat{P}\gamma^\dagger_{\bf
k\sigma}|\Phi\rangle,
\label{eq:phik}
\ee
where
\be
\gamma^\dagger_{\bf k\uparrow} = u_{\bf k}
c^\dagger_{\bf k\uparrow}-\hat{S}v_{\bf k}c_{-\bf k\downarrow}
\label{eq:gammak}
\ee
and the operator $\hat{S}$ creates a ground-state pair. The procedure is
entirely analogous to the Hartree-Fock-BCS theory where the
ground state is determined by the criterion that all single-Fermion
excitations have positive energy.  Within this theory, the
excitations in Eq.~\ref{eq:phik} are the low energy single-Fermion
excitations, by Koopman's theorem, which may be demonstrated in this
case by
the same method as in normal Hartree-Fock.

We note that the theory so far, and its manipulations, are only
correct because the Hamiltonian conserves particle number, so
that we do not need to consider coherence between states with different
particle numbers.   In order to specify that the number  of
electrons is correct we may  simply fix average occupancies at the
appropriate values, $x$ for the empty state and $\frac12(1-x)$ for the singly
occupied ones of given spin; and then we proceed with Gutzwiller
approximation based on those occupancies.  But if, as in tunneling, we
need to
add or subtract electrons, we must follow Laughlin~\cite{Laughlin} and
introduce a fugacity factor $Z$ for electron pairs.  This is easily
computed
by noting that the ratio of these two occupancies in the original
product function is $(1-x)/(1+x)$, so to get the correct
occupancies we must correct the normalization by $(2x/(1+x))^{1/2}$.  Hence for a pair of holes
the fugacity factor is
\be
Z = \frac{2x}{(1+x)}.
\label{Z}
\ee
The resulting wave function, in the form
given by Laughlin~\cite{Laughlin}, is Eq. \ref{bcs} multiplied by a
factor
$Z^{-(n_\uparrow+n_\downarrow)}$.  (We choose for perspicuity to
express $Z$ as the fugacity of holes rather than electrons; the
choice is arbitrary).

This wave function may be rewritten in a form which demonstrates
the effect of $Z$ more clearly.  In each factor
$(u_{\bf k} +v_{\bf k}c^\dagger_{\bf
k\uparrow}c^\dagger_{-k\downarrow})$
the $v_{\bf k}$ factor
creates a pair of electrons, or conversely the $u_{\bf k}$ factor can be
taken as creating a pair of holes; thus in any component of the
wave function which contains $u_{\bf k}$, we can insert a corresponding
factor $Z$.  Thus
instead of Eq.~\ref{bcs} we could write
\be
|\Phi^\prime\rangle = \prod_{\bf k}
(Zu_{\bf k}+v_{\bf k}c^\dagger_{\bf k\uparrow}c^\dagger_{-\bf
k\downarrow})|0\rangle,
\label{eq:bcs2}
\ee
and then we have absorbed the fugacity factor into $\Phi$.  But
the individual factors in Eq.~\ref{eq:bcs2} are not normalized.  To
define the appropriate quasiparticle excitations as in
Eq.~\ref{eq:gammak},
they must be normalized and thus, finally, the appropriate product
function
must be written
\be
|\Phi^{\prime\prime}\rangle = \prod_{\bf k}
{{(Zu_{\bf k}+v_{\bf k}c^\dagger_{\bf k\uparrow}c^\dagger_{-\bf
k\downarrow})}\over{\sqrt
{u_{\bf k}^2Z^2+
v_{\bf k}^2}}}|0\rangle
\label{eq:bcs3}
\ee

The individual factors define the product $\gamma_{\bf k}\gamma_{-\bf
k}$,
so that the individual gamma contains the normalizing factor $(u^2Z^2 +
v^2)^{\frac14}$.  Note that the inclusion of the $Z$ factors nicely
leads
to the $Z$ renormalization of the order parameter, the superfluid
density, and the kinetic
energy~\cite{Zhang}.  $Z$ plays a role which can be described as
the amplitude of the
hole-pair wave function -- or at least the relative amplitude of
the part of the pair function which is holes as opposed to spins.

We would like to emphasize that the wave function (9) is
completely identical to that used in previous papers and that no results as
to quasiparticle energies or ground state averages are at all
modified.   The quantities $u$ and $v$ define the starting wave function
which is to be projected and its populations modified.  We can think of the
fugacity factor as part of the Gutzwiller projector, if we like;
and like the projector, it does not commute with the fermion
operators.  Thus when we need to express the excitations in terms of
real Fermions added to the system -- outside the projector -- we must use
a modified set of $u$ and $v$.  But the real Fermions  predominantly go
in coherently as single quasiparticles, except for the relatively
small term for holes coming from the fluctuations of the opposite-spin
occupancy.

We want to present the simplest possible calculation, since the
effect in question is a qualitative one. To this end we set
$\mathrm{e}^{iS} = 1$, which involves errors of the order $J/t$ which vary smoothly with
energy -- we are neglecting virtual double occupancy, and will
therefore overestimate the asymmetry somewhat.

Ignoring $\mathrm{e}^{iS}$, the quasiparticle wave function can be created
from the unprojected product function $|\Phi^{\prime\prime}\rangle$
by either of the operators $\hat{P}c^\dagger_{\bf k\uparrow}$ or $\hat{P}c_{-\bf k\downarrow}$.
The former creates it with relative
amplitude $uZ/(Z^2u^2+v^2)^{\frac14}$, the latter with relative
amplitude
$v/(Z^2u^2+v^2)^{\frac14}$.
But what tunnels in from the metal is not a projected
quasiparticle but a
real one. We can think of the STM point as instantaneously
depositing
a particle or hole into the Wannier function at the Cu atom under
the
point, and we then Fourier resolve the Wannier function amplitude
at
a time $t$ = 0 into excitations in the superconductor.
Thus the operators which we must consider, acting on our assumed
ground state, are $c^\dagger_i\hat{P}$ and $c_i \hat{P}$.
For present purposes we can discuss
only the site operators, later Fourier transforming to get the
momentum space ones.

Consider $c^\dagger_{i\sigma}\hat{P}$.
This may be divided into its two parts belonging to
the forbidden and allowed subspaces:
\be
c^\dagger_{i\sigma}\hat{P}=(1-\hat{P}) c^\dagger_{i\sigma}\hat{P} +
\hat{P}c^\dagger_{i\sigma}\hat{P}.
\label{eq:c}
\ee

The first term contains only excitations with energy larger than
the
Hubbard $U$ and does not concern us. The second term is finite
only $x$
of the time; it requires that the site $i$ contains an electron
in the
ground state function $\hat{P} |\Phi^{\prime\prime}\rangle$.
Thus the probability that it creates an
excitation in the allowed manifold is multiplied by $x$.  But it
will
be important to note that because $\hat{P}c^\dagger \hat{P}=\hat{P}c^\dagger$,
obviously, the part of the hole that doesn't go into the upper Hubbard
band  creates
only single-quasiparticle excitations in this approximation, thus has
only a coherent spectrum. Now consider $c\hat{P}$. This automatically goes
into
the allowed manifold, but $c\hat{P}|\Phi^{\prime \prime}\rangle$
is not exactly the same as the
quasiparticle function $\hat{P} c |\Phi^{\prime\prime}\rangle$
because the latter can contain
components where $|\Phi^{\prime\prime}\rangle$ has $n_i = 2$
with probability
$(1 - x^2)/4$, while these do not appear in
$c \hat{P} |\Phi^{\prime\prime}\rangle$.
To adjust the
normalization, note that
\be
c_\uparrow \hat{P} = c_\uparrow (1-n_{\uparrow}n_{\downarrow}) = \hat{P}
(1-n_\downarrow)c_\uparrow,
\label{eq:comm}
\ee
where the site index has been dropped. The average factor
reducing the quasiparticle function is
\be
c\hat{P} = \hat{P}c(1- < n_\downarrow >) = {{(1+x)}\over {2}} \hat{P} c.
\label{comm2}
\ee

Thus the ratio of the normalization factors for the electron vs
the
hole spectra is $g_t =Z=2x/(1 + x)$. But in this case there is an
incoherent spectrum, caused by the three-Fermion operator
$[n_\downarrow-<n_\downarrow>]c^\dagger$;
we do not believe this is a large effect, but it may
cause features in the hole spectrum, particularly in the
neighborhood
of $\Delta$ added to the magnetic resonance energy.

For each ${\bf k}$ and spin, except for the rather small term mentioned
in the last paragraph, there is only one quasiparticle wave function
that appears in this approximation, that obtained by Gutzwiller
projecting the suitable BCS product function.  Most of the
amplitude
is ``coherent",  a conclusion quite different from that of
Wen~\cite{Wen}.

At first sight one would think that the ratio of the tunneling
conductivity for the sign of  voltage $V$ such as to inject a
hole -- external electrode positive -- to that with the opposite sign
would be just $2x/(1+x)$. But actually, quasiparticles are not
pure electrons or holes but mixtures
of the two, and precisely at the gap energy they are equal
mixtures, so that at the gap energy the tunneling
conductivity for $+V$ and $-V$ should be equal. The relevant
tunnel current can flow either in the form of right-moving holes or
left-moving electrons, and in the superconductor it is an equal
coherent mixture of the two. But it is important to realize that
for a given sign of voltage the two states which are coherent have
actually the same charge, so that the hole is accompanied by a
ground-state pair.

Our calculation follows the method of Tinkham~\cite{Tinkham}.  As
he points out (following in this Cohen, Falicov and
Phillips~\cite{Cohen})
there are two quasiparticle channels with the same energy, with
$(\epsilon_{\bf k} - E_f)$ positive and negative -- electron-like and
hole-like respectively.
The $u_{\bf k}$ and $v_{\bf k}$ of the hole-like channel exchange their
values,
so that $u$(hole-like) =$v$(electron-like), and vice versa. Thus the
conductance is
symmetric in the exchange of $u$ and $v$, but NOT in the exchange of
holes and electrons, in contrast to the BCS case.

For electrons,  the current is first of all reduced by the
projection factor $Z$ relative to that for the holes.  Then the
amplitude
for a given channel is just the effective $u$ for that channel, and the
current its square; we get, taking into account the
renormalization factor,  that the tunneling density of states for
electrons is
\be
N_{e}(E,\Delta)={{d\epsilon}\over{dE}} Z \left( {{u^2}\over{\sqrt
{u^2+v^2Z^2}}}
+ {{v^2}\over{\sqrt{v^2+u^2Z^2}}} \right),
\label{Ne}
\ee
where
$u^2=\frac12[1+\epsilon/E]$
and $v^2=\frac12[1-\epsilon/E]$.

For holes,  we have no projection factor $Z$, but the hole
amplitude contains the factor $Z$ which can be thought of as the
magnitude of the pair wave function.  This satisfies the physical
requirement that the coherent amplitude for holes and electrons
must be the same at least at the same energy, and is also necessary
for equilibrium. But the renormalization factor is not identical
except
at $\epsilon=0$, $E=\Delta$ and rises as $E\rightarrow\infty$ to
$1/Z$:

\be
N_{h}(E,\Delta)={{d\epsilon}\over{dE}}Z
\left( {{v^2}\over{\sqrt{u^2+v^2 Z^2}}} +
{{u^2}\over{\sqrt{v^2+u^2 Z^2}}} \right ).
\label{Nh}
\ee
Equations \ref{Ne} and \ref{Nh} show that the coherence peaks at
$\epsilon = 0$ are
identical as predicted,  but the ratio of the $E\rightarrow
\infty$
asymptotes is $Z$,
as expected from simple considerations.

\bfig[h]   % Fig 1
\incl[width=8cm]{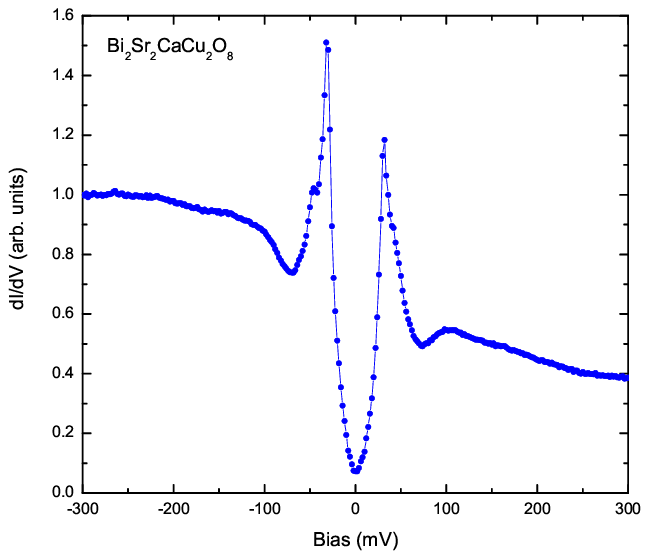}
\caption{\label{Pan} Tunneling conductance vs voltage for an optimally
doped sample of BSCCO [data from S H Pan (unpublished)].  The
cleanest data has been selected from among several samples.
Observations of J C Davis' group, and earlier work of
Fischer~\cite{Fischer},
are similar in their general course.
}
\efig
Finally it is necessary  to take into account that we have a
$d$-wave gap which means that the tunnel current must be integrated over
the gap distribution
\be
{\cal P}(\Delta)d\Delta = \frac{d\Delta}{\sqrt{1-(\frac{\Delta}{\Delta_0})^2}},
\label{Pgap}
\ee
with $\Delta_0$ the gap amplitude.

The differential conductance for injection of electrons or holes is then
\be
G(E)_{e,h} = \int_0^{\Delta_0} d\Delta \;\; N_{e,h}(E,\Delta){\cal P}(\Delta) d\Delta,
\label{dIdV}
\ee
with $N_{e,h}$ given by Eqs. \ref{Ne} and \ref{Nh}, respectively.  
For numerical convergence, we add a weak imaginary part to the gap
to simulate broadening, i.e.
$\Delta\rightarrow (1+\mathrm{i}\eta)\Delta$ with $\eta = 2\times 10^{-3}$.

We have approximated the Fermi Surface by a circle and normalized
the maximum gap to 1.  The result for the symmetric conductivity in
the BCS case has often been displayed, and involves an elliptic
function in its analytic form; there is a logarithmic peak at
$\Delta_0$,
and a linear slope at $E$ near zero. These shapes will appear in the
region of the coherence peak and below in the Gutzwiller case, since the
asymmetry  in Eqs. \ref{Ne} and \ref{Nh} is of greater than linear order
in $u^2-v^2=\epsilon/E$.  But the renormalizations become appreciable
even near $E=\Delta_0$, and especially on the hole side can rise
to dominate the peak for the underdoped case.  The asymmetry has a
pronounced
upward cusp at $E=\Delta_0$.
\bfig[h]   % Fig 2
\incl[width=7cm]{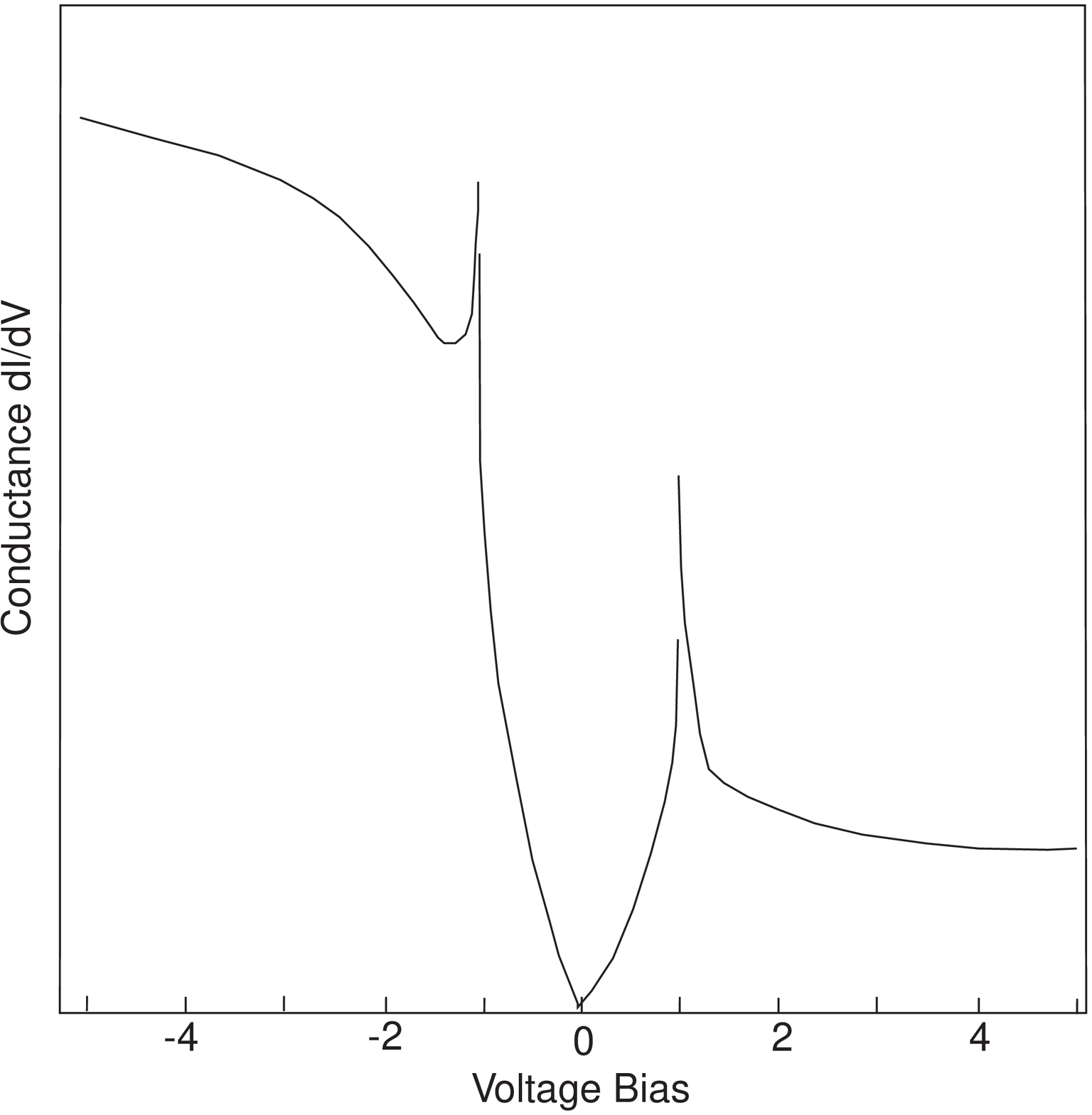}
\caption{\label{compGV} Computed curve of conductance vs voltage for $Z$
= 0.2,
which is too underdoped to fit Fig. \ref{Pan} (in which we estimate
$Z\simeq$ 0.3) but the gap
distribution is realistic to give a reasonable amplitude of
coherence peak.
}
\efig
\bfig[h]   % Fig 3
\incl[width=9cm]{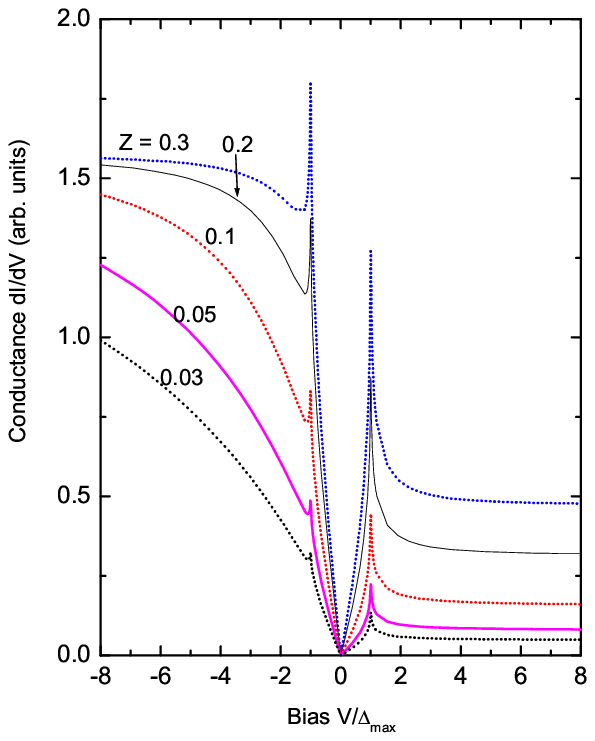}
\caption{\label{GV} Computed curves for a sequence of $Z$'s to
demonstrate the variation with doping and to allow 
qualitative extrapolation to $Z$ = 0.  To achieve convergence, a complex gap 
$\Delta(1+\mathrm{i}\eta)$ ($\eta = 2\times10^{-3}$) is used 
in the distribution Eq. \ref{Pgap}.  The curves bear a rough 
resemblance to extremely underdoped experimental results.
}
\efig
It has been hard to find an accurate curve for comparison of the
predicted asymmetry with the published data. The best published
data we can estimate from, for instance, Pan \etal~\cite{Pan2}, seems in surprisingly
good agreement, especially when we realize that it is likely to
contain traces of the interaction with the magnetic resonance of
Keimer~\cite{Keimer}

The best curve from Pan's data is shown in Fig \ref{Pan}, which
refers to a $\mathrm{Bi_2Sr_2CaCu_2O_8}$ (BSCCO) sample at optimal
doping.  It
should be noted that it is extremely difficult experimentally to
make the conditions exactly identical for + and - $V$, and it is
therefore likely
that the very small asymmetry in the coherence peak structure is an
experimental artifact, and is actually absent.  Ignoring that,
the fit to the general course of the asymmetry is remarkable. Other
data not shown here confirm the general behavior with doping as well.

A better estimate of the experimental curve would take into
account that the band is much flatter at the gap antinodes near 0, so
that the maximum of the gap should be more strongly emphasized.  An
estimate of such an effect can be obtained by adding to ${\cal
P}(\Delta)$ in Eq. \ref{Pgap}
a delta-function of unity at $\Delta_0$; Fig. \ref{compGV} shows
the result of such an
exercise, which seems to reproduce the experimental curve of Pan
at optimal doping fairly well.

Another somewhat speculative exercise is to continue $Z$ towards 0,
which gives us a conjectured tunneling spectrum for the pure RVB
phase, which is our model for the pseudogap state, at least at
higher temperatures.  In Fig. \ref{GV} we show how the variation with
doping
goes. In the limit as $Z\rightarrow 0$, the current (almost all on
the hole side) does not extrapolate to an asymptote but continues to
rise linearly at high voltage.  The ratio of the asymptotes on the two
sides is $1/Z$, an observation which seems to accord with most
estimates of doping percentages.

The curves which represent regions the experimentalists think are
quite underdoped differ from higher dopings in that the
symmetrical parts of the curve  extend only to rather low voltages, and
the coherence peaks are suppressed.  The higher-voltage conductivity
seems almost completely composed of the asymmetric ``hump"
behavior and to be dominantly on the hole side.  In the same regions, a
characteristic ``$4\times 4$" density wave is observed.  In a
forthcoming paper we will suggest a mechanism that might relatively suppress
the gap antinodes.

We would like to acknowledge extensive discussions with J. C.
Davis, and the stimulus of his brilliant analysis of his data, some of
it unpublished. S. H. Pan shared useful unpublished data with us,
and \O. Fischer and L. Greene have also shown us some of their results.
Above all, we would liike to acknowledge the diligent assistance of 
Mohit Randeria, Nandini Trivedi and Fu-Chun Zhang in correcting our 
misconceptions and clarifying our thinking on these issues, without 
which effort this paper could not have been written.

\end{document}